\documentclass[prl,epsfig,floats,twocolumn,superscriptaddress,amssymb,floatfix,showpacs]{revtex4}
\usepackage{graphicx}% Include figure files
\usepackage{dcolumn}% Align table columns on decimal point
\usepackage{bm}% bold math
\begin{document}

\title{Non-contact rack and pinion powered by the lateral Casimir force}

\author{Arash Ashourvan}
\affiliation{Institute for Advanced Studies in Basic Sciences,
Zanjan, 45195-1159, Iran}

\author{MirFaez Miri }
\email{miri@iasbs.ac.ir} \affiliation{Institute for Advanced Studies
in Basic Sciences, Zanjan, 45195-1159, Iran}

\author{Ramin Golestanian}
\email{r.golestanian@sheffield.ac.uk} \affiliation{Department of
Physics and Astronomy, University of Sheffield, Sheffield S3 7RH,
UK}

\date{\today}
\begin{abstract}
The lateral Casimir force is employed to propose a design for a
potentially wear-proof rack and pinion with no contact, which can be
miniaturized to nano-scale. The robustness of the design is studied
by exploring the relation between the pinion velocity and the rack
velocity in the different domains of the parameter space. The
effects of friction and added external load are also examined. It is
shown that the device can hold up extremely high velocities, unlike
what the general perception of the Casimir force as a weak
interaction might suggest.
\end{abstract}


%{42.50.Lc}{Quantum fluctuations, quantum noise, and quantum jumps}
%{03.70.+k}{Theory of quantized fields}
%{42.50.-p}{Quantum optics}

\maketitle

%%%%%%%%%%%%%%%%

With the emergence of the new generation of miniaturized mechanical
devices such as micro- and nano-electromechanical systems (NEMS), we
have been witnessing a paradigm change in the technical problems
involved in making machines and the strategies needed to resolve
them. In particular, tribological interactions, i.e. friction,
adhesion, and wear, appear to pose new challenges at small length
scales \cite{Carpick}. The abundance of surfaces in contact in small
devices and high friction are problematic, and the traditional use
of lubricants does not work because they become excessively viscous
when made into molecular layers \cite{Lubric}. Devices with sliding
surfaces in contact are known to wear out too rapidly \cite{Wear},
and it seems that strategies to minimize contact between the
surfaces are needed to help make them more durable. The presence of
short-ranged attractive dispersion or Casimir forces can cause tiny
elements in small devices to stick together and bring the devices to
stop \cite{Buks}, and a line of ongoing active research is focused
on devising novel techniques to avoid such effects
\cite{recent-stra1,recent-stra2}. In light of these technical
difficulties, it seems desirable to have novel designs for
mechanical devices that can operate without physical contact between
their parts.

As a key interaction at nano-scale, Casimir force can be harnessed
and used in small devices, as demonstrated by Capasso and
collaborators who developed an actuator powered by the normal
Casimir force between a flat plate and a sphere
\cite{Capasso1,Capasso2}. To avoid the limited applicability that
the parallel-plate geometry might offer, one can make the two
surfaces corrugated and take advantage of a lateral component to the
Casimir force, as has been recently proposed \cite{golestan97} and
indeed verified experimentally \cite{Mohideen}. The lateral Casimir
force between corrugated surfaces, provides a possibility for
friction-less transduction of lateral forces in nano-mechanical
devices without any physical contact between them. The coupling
between two surfaces via the quantum vacuum is realized by a term
proportional to the sinus of the phase difference between them and
is a macroscopic manifestation of quantum coherence, which is
reminiscent of the Josephson coupling between superconductors
\cite{golestan97,golestan98}. Because the coupling is nonlinear, its
mechanical response could involve oscillatory and unstable behavior
similar to the what is observed in Ref. \cite{Capasso2} for the case
a device powered by the normal Casimir force.

\begin{figure}[b]
\includegraphics[width=0.9\columnwidth]{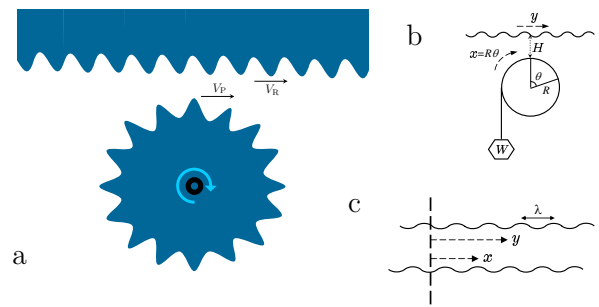}
\caption{(a) The rack and pinion with no contact. (b) The schematics
of the rack and pinion with an external load $W$. (c) Two corrugated
plates with lateral shift $x-y$. The equilibrium position is at
$x=y$.} \label{fig:gear}
\end{figure}

Here we make use of the lateral Casimir force to design a nano-scale
rack and pinion without intermeshing, as shown in Fig.
\ref{fig:gear}a. Our system consists of a corrugated plate (rack)
and a corrugated cylinder (pinion) that are kept at a distance from
each other. The pinion could be subject to external load (see Fig.
\ref{fig:gear}b), and could experience friction when rotating around
its axis. For uniform motion of the rack with a velocity $V_{\rm
R}$, we find that the ``contact'' pinion velocity $V_{\rm P}$ (see
Fig. \ref{fig:gear}a) is locked in to the rack velocity for
sufficiently small values of $V_{\rm R}$, and that there is a
threshold rack velocity at which the pinion undergoes a {\em
skipping transition} where the pinion can no longer hold the cogs in
perfect registry with the rack. In the skipping regime, we find that
the average pinion velocity can be both positive and negative (with
$V_{\rm R}>0$) depending on the initial phase mismatch between the
corrugations. The effects of the external load and friction are
considered and the regions in the parameter space where the pinion
can do work against the load are determined. These studies could
help us examine the feasibility and efficiency of such a design as a
mechanical transducer.

%\section{Theoretical Formulation}

When two sinusoidally modulated plates with identical wavelengths of
$\lambda$ are shifted with respect to each other by a length $x-y$
(see Fig. \ref{fig:gear}c), the lower plate experience the lateral
Casimir force
\begin{math}
F_{\rm lateral}= -F \sin\left[\frac{2 \pi}{\lambda}
(x-y)\right]
%,\label{F-lat}
\end{math}
\cite{golestan97,golestan98} where the amplitude $F$ depends on the
mean separation of the plates and the amplitude of corrugations
\cite{Mohideen,Emig-etal-2001,Reynaud}. The lateral Casimir force
introduces a net torque on the pinion, which plays the central role
in the equation of motion for the coordinate $x=R \theta$ (Fig.
\ref{fig:gear}b). The equation of motion reads
\begin{equation}
\frac{I}{R} \frac{d^2 x}{d t^2}=-R F \sin\left[\frac{2 \pi}{\lambda}
(x-y)\right]-\frac{\zeta}{R}\frac{d x}{d t}-R W, \label{master1}
\end{equation}
where $I$ is the moment of inertia of the pinion about its major
axis, $\zeta$ is the rotational friction coefficient, and $W$ is an
external load against which the pinion should do work.

We focus on the uniform motion of the rack with a velocity $V_{\rm
R}$, i.e. $y=V_{\rm R} t$, although similar analyses can be
performed for other types of motion such as vibrating \cite{AFR-unp}
or undulating \cite{Emig-unp} racks, in both of which cases the
motion can be rectified. The nonlinear equation \ref{master1} can be
better studied in the phase plane $\left(u \equiv 2 \pi
(x-y)/\lambda, v=\dot{u} \right)$, where it reads $\dot{u}=v$,
$\dot{v}=-\sin u-\epsilon (v-v_0)-W/F$. In the above equations, we
have measured the time in units of $T=\sqrt{I \lambda /(2 \pi F
R^2)}$, and have defined the dimensionless parameter $\epsilon= T
\zeta /I$, which is a measure of the relative importance of friction
in the system. The parameter $v_0=\dot{u}_0=2\pi(\dot{x}_0- V_{\rm
R} T)/\lambda$ is the initial value for $v$. A key velocity scale is
given by
\begin{equation}
V_{\rm S}=\frac{\lambda}{2 \pi T}=\left(\frac{F \lambda R^2}{2 \pi
I}\right)^{1/2},\label{VX-def}
\end{equation}
which corresponds to the velocity at which the kinetic energy of the
rotating pinion is of the order of the work done by the lateral
Casimir force upon displacement by one tooth. As we will see below,
the quantity $F \lambda$ can be considered as the effective ``bond
strength'' of the coupling between the pinion and rack, so that a
``bound state'' between them can only tolerate pinion kinetic
energies of this order or magnitude, which means that Eq.
\ref{VX-def} gives the {\em skipping velocity}. For simplicity, we
only consider the case of pinions that are initially at rest
throughout this paper ($\dot{x}_0=0$), which means $v_0=-V_{\rm
R}/V_{\rm S}$, although the formulation can be readily used to study
other initial conditions as well.

\paragraph{No dissipation.}

In the absence of dissipation and load, Eq. \ref{master1} is
identical to the celebrated nonlinear plane pendulum problem
\cite{Guck}. It is well known that for this system, the ``energy''
\begin{math}
h=\frac{1}{2} v^2+ 1- \cos u=\frac{1}{2} \left(\frac{V_{\rm
R}}{V_{\rm S}}\right)^2+ 1- \cos u_0,\label{h-def}
\end{math}
is a constant of motion, and there are two families of periodic
orbits, corresponding to rotations ($h>2$) and oscillations ($h<2$).
For $0<h<2$, the system is oscillatory with a period of $4
K(\sqrt{h/2})$, where $K(m)=\int_0^{\pi/2} d\theta (1-m^2 \sin^2
\theta)^{-1/2}$ is the complete elliptic integral of the first kind
\cite{GradRyz}. In this case $|u|\leqslant |\cos^{-1}(1-h)|
\leqslant \pi $, which means that the distance $|x-y|$ between the
teeth of the rack and pinion does not exceed $\lambda/2$. Thus the
pinion is locked-in with the rack and will have a forward motion
with superimposed oscillations (jerks). In the case of $h>2$, the
system is not locked in any more and depending on the initial
conditions it can have different behaviors.

\begin{figure}[tbp]
\includegraphics[width=0.99\columnwidth]{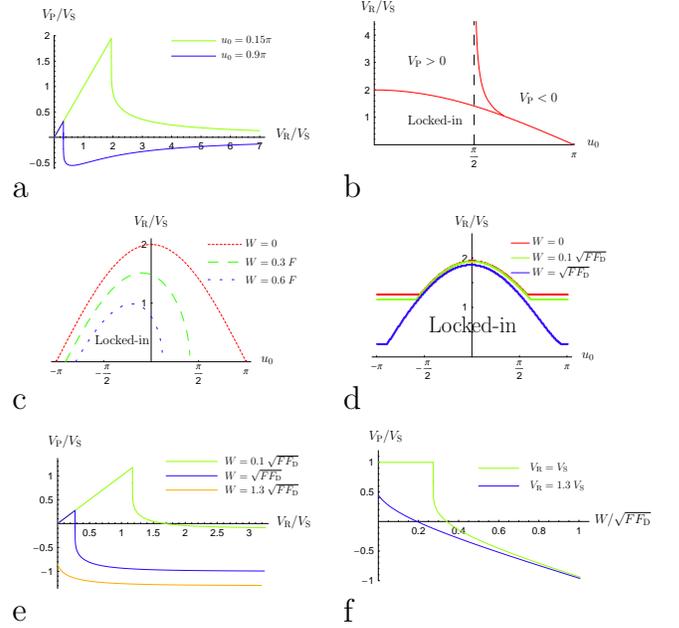}
\caption{(a) Pinion velocity versus rack velocity in the absence of
dissipation and external load. (b) Domains of positive and negative
pinion velocities as well as the phase boundary for the skipping
transition. Only half of the plot is shown due to $u_0 \to -u_0$
symmetry in this case. (c) Phase boundary for the skipping
transition, in the absence of dissipation, as a function of the
external load. (d) Phase boundary for the skipping transition, in
the presence of weak dissipation, as a function of the external load
for $\sqrt{F_{\rm D}/F}=0.05$. (e) Pinion velocity versus rack
velocity in the presence of weak dissipation and external load. (f)
Force--velocity response of the rack and pinion could be of two
general forms depending on the value of the rack velocity.}
\label{fig:VRVP-combo}
\end{figure}

The value of $h$, which determines the behavior of the system,
depends on $V_{\rm R}$ and $u_0$. For $V_{\rm R}<V_{\rm S} \sqrt{2
(1+\cos u_0)}$ the rack and pinion are geared up and we have $V_{\rm
P}=V_{\rm R}$ for the jerk--averaged pinion velocity. Higher rack
velocities $V_{\rm R}>V_{\rm S} \sqrt{2 (1+\cos u_0)}$ cause the
system to go to the ``rotation'' phase of the equivalent pendulum
problem, which means that the pinion skips teeth with respect to the
rack, but will still have a net average velocity that can be
calculated as
\begin{equation}
V_{\rm P}=V_{\rm R}-\frac{\pi V_{\rm S}}{\sqrt{{2}/{h}}\;
K\left(\sqrt{{2}/{h}}\right)},\label{VP-def}
\end{equation}
%where $h$ is given by Eq. \ref{h-def}.
In Fig. \ref{fig:VRVP-combo}a, the pinion velocity is plotted as a
function of the rack velocity for two values of the initial phase
mismatch.
%$u_0=0.15 \pi$ and $u_0=0.9 \pi$.
The pinion velocity rises linearly with the rack
velocity initially and then drops abruptly at the skipping
transition. At large rack velocities $V_{\rm R} \gg V_{\rm S}$, the
pinion velocity has an asymptotic form
\begin{math}
V_{\rm P}=\cos u_0 {V_{\rm S}^2}/{V_{\rm R}}+\cdots,
\end{math}
which shows that it vanishes at infinity, and that the decay can be
both from below (for $u_0> \pi/2$) and above (for $u_0<\pi/2$). An
intriguing feature here is the possibility of dropping into negative
pinion velocities---the {\em reverse gear}---after the skipping
transition, as Fig. \ref{fig:VRVP-combo}a shows. Figure
\ref{fig:VRVP-combo}b delineates the different domains of phase
lock-in, forward motion, and reverse motion, in the space of the
rack velocity and the phase mismatch. One can see that for $0 < u_0
< \pi/2$ the average velocity is always positive, while for $\pi/2 <
u_0 < \pi$ the pinion velocity can be switched from positive to
negative by increasing $V_{\rm R}/V_{\rm S}$. This can be achieved
by either speeding up the rack or decreasing the skipping velocity
by increasing the separation between the rack and pinion (see
below).

\paragraph{The effect of external load.}

In the absence of dissipation, Eq. \ref{master1} has
$h'=v^2/2+1-\cos u+W u/F$ as the constant of motion.
%The center and saddle fixed points are located at
%$\left(-\sin^{-1}(W/F),0 \right)$ and
%$\left(-\pi+\sin^{-1}(W/F),0\right)$, respectively.
Two classes of motion are separated by the saddle-loop of energy
${h'}_{s}=1+\sqrt{1-(W/F)^2}-[\pi-\sin^{-1}(W/F)] (W/F)$, leading to
rotations for $h'>{h'}_{s}$ and oscillations for $h'<{h'}_{s}$, in
the equivalent pendulum problem. Figure \ref{fig:VRVP-combo}c shows
the domains for two different regimes as well as the boundary for
the skipping transition as a function of the rack velocity and the
initial phase mismatch. As the load is increased, the two fixed
point approach each other thereby narrowing down the locked-in
region, until at $W=F$ it disappears when they fully merge.
Therefore, $W<F$ is a necessary condition to have positive rack
velocities.
%Note that the entire skipping region corresponds to
%$V_{\rm P} <0$ because the constant of motion prohibits unlimited
%buildup of positive values of $u$ for any non-vanishing $W/F$.

\paragraph{Weak dissipation.}

The dynamical system described by Eq. \ref{master1} is not
integrable in the presence of the dissipation term, as energy is not
conserved. If the dissipation is weak, such that $\zeta V_{\rm
R}/R^2+W < F$, we can use the Melnikov method \cite{Guck,melnikov}
to study the perturbed phase portrait of the system. In this regime,
the force scale
\begin{math}
F_{\rm D}=\frac{\zeta^2 \lambda}{2 \pi I R^2}
%,\label{FD-def}
\end{math}
seems to play an important role. (Note that ${\zeta V_{\rm
S}}/{R^2}=\sqrt{F F_{\rm D}}$.)

Similar to the dissipation-free case, the system develops the two
classes of commensurate and incommensurate motion transduction
separated by a skipping transition boundary. Figure
\ref{fig:VRVP-combo}d shows an example of this phase diagram that
has been calculated numerically, for different values of the
external load. The phase boundary consists of horizontal
($u_0$-independent) parts that are met by a bell-shaped central
part. In the part that does not depend on $u_0$, we find that for
$\zeta V_{\rm R}/R^2< \frac{4}{\pi} \sqrt{F F_{\rm D}}-W$ the system
is in the locked-in phase and we have $V_{\rm P}=V_{\rm R}$. As the
rack velocity is increased above this limit, the system undergoes a
skipping transition. The pinion velocity for $\zeta V_{\rm R}/R^2>
\frac{4}{\pi} \sqrt{F F_{\rm D}}-W$ can be found from Eq.
\ref{VP-def}, with a value $h=h_m$ that is a solution to this
equation
\begin{math}
V_{\rm R}+W R^2/\zeta=\frac{4}{\pi} V_{\rm S}
\;\sqrt{h_m/2}\;E\left(\sqrt{{2}/{h_m}}\right)
%,\label{hm-def}
\end{math}
with $E(m)=\int_0^{\pi/2} d\theta \sqrt{1-m^2 \sin^2 \theta}$ being
the complete elliptic integral of the second kind \cite{GradRyz}.
Figure \ref{fig:VRVP-combo}e shows the pinion velocity as a function
of the rack velocity for different values of the external load. The
general feature of a drastic drop in the pinion velocity after the
skipping transition is observed, and the asymptotic behavior
\begin{math}
V_{\rm P}=-W R^2/\zeta+\frac{1}{2} {V_{\rm S}^4}/{V_{\rm
R}^3}+\cdots,
\end{math}
at large rack velocities shows a complete decoupling in the system
at infinitely large $V_{\rm R}$. The response of the system in the
bell-shaped region of the phase diagram of Fig.
\ref{fig:VRVP-combo}d leads to similar results and can be obtained
numerically. Note that the Melnikov approximation breaks down for
large rack velocities and we need a complementary approach to
examine the behavior of the system in that limit (see below).

The force--velocity response of the system in the $u_0$-independent
region can also be extracted from this result. Figure
\ref{fig:VRVP-combo}f shows the pinion velocity as a function the
external load for two values of the rack velocity. For $V_{\rm R}<
\frac{4}{\pi} V_{\rm S}$, the pinion velocity is independent of the
load for $W< \frac{4}{\pi} \sqrt{F F_{\rm D}}-\zeta V_{\rm R}/R^2$,
and drops drastically after the onset of skipping, whereas the
decrease upon introduction of load starts from the beginning when
$V_{\rm R}> \frac{4}{\pi} V_{\rm S}$. One can identify the load at
which the pinion velocity vanishes as the {\em stall force}.

\paragraph{Strong dissipation.}
The Melnikov method fails if $\zeta V_{\rm R}/R^2+W \gtrsim F$, but
in this limit we can proceed by neglecting the acceleration term
$\dot{v}=\ddot{u}$ in Eq. \ref{master1}. This allows us to solve the
equation in closed form, which yields $u(t)=2 \tan^{-1}
\left[\left(\frac{F}{\zeta V_{\rm R}/R^2+W}\right)
-\sqrt{1-\left(\frac{F}{\zeta V_{\rm R}/R^2+W}\right)^2}\; \tan
\left(\frac{\pi t}{\tau}\right)\right]$, where
$\tau=\lambda/\sqrt{(V_{\rm R}+W R^2/\zeta)^2-(F R^2/\zeta)^2}$. The
explicit expression for $u$ can be used to calculate the
time-averaged pinion velocity
\begin{math}
V_{\rm P}=V_{\rm R}-\sqrt{(V_{\rm R}+W R^2/\zeta)^2-(F
R^2/\zeta)^2}
%,\label{VP-toofric-def}
\end{math}
which produces very similar curves to those plotted in Fig.
\ref{fig:VRVP-combo}e above the skipping transition (using the
Melnikov method). The stall force (load) in this limit can be found
as $W_{s}=F \left[\sqrt{1+\left({\zeta V_{\rm R}}/{F
R^2}\right)^2}-\left({\zeta V_{\rm R}}/{F R^2}\right)\right]$.
%Interestingly, Eq. \ref{VP-toofric-def} gives a limiting value of
%$V_{\rm P}=-W R^2/\zeta$ at large values of rack velocity, in
%agreement with the asymptotic behavior suggested by the Melnikov
%approach.

%\section{Concluding Remarks}

\begin{figure}
\includegraphics[width=0.85\columnwidth]{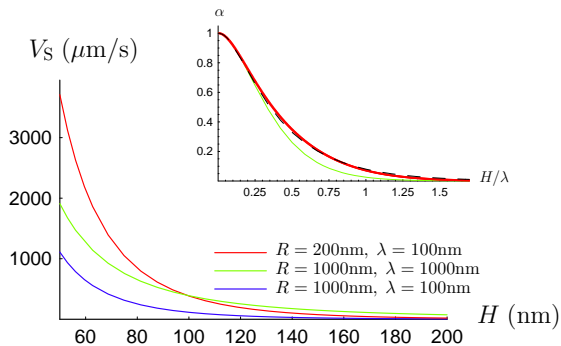}
\caption{Skipping velocity as a function of the gap size for perfect
metallic boundaries, corresponding to $\rho=19.3\; {\rm gr}/{\rm
cm}^3$ (gold) and $a_1=a_2=10$ nm, for different values of radius
and corrugation wavelength. Inset: the parameter $\alpha$ introduced
in the text as a function of $H/\lambda$ for perfect metals (red
dots) compared to the pairwise summation approximation (green solid
line). The dashed-line shows the empirical formula presented in the
text.} \label{fig:derjaguin}
\end{figure}

To make a stronger link to the experiments, we need to quantify the
amplitude of the lateral Casimir force $F$. For a cylinder of radius
$R$ located at a (nearest) distance $H$ from a plate (see Fig.
\ref{fig:gear}b), the Casimir interaction can be calculated from the
corresponding interaction between two parallel plates, using the
Proximity Force Approximation (PFA) \cite{prox}. For the normal
Casimir force between perfect metals, it has been recently shown
that this approximation works surprisingly well for $H \lesssim R$
\cite{Kardar-cyl}. Using PFA for the lateral Casimir force between a
pinion of length $L$ and corrugation amplitude $a_1$ and a rack of
corrugation amplitude $a_2$ with $a_1, a_2 \ll H$, we find
\begin{equation}
F=\left(\frac{7 \pi^4 \sqrt{2}}{3072}\right) \frac{\hbar c a_1 a_2 L
R^{1/2}}{\lambda H^{9/2}}
\;\alpha\left(\frac{H}{\lambda}\right),\label{F-res-1}
\end{equation}
to the leading order, where
\begin{math}
\alpha=\frac{3072}{7 \pi^3} \int_1^\infty \frac{d t}{t^5
\sqrt{t-1}}J\left(\frac{H}{\lambda}t\right)
\end{math}
with $J$ being the (``Josephson--'') coupling function presented in
Ref. \cite{Emig-etal-2001}. The inset of Fig. \ref{fig:derjaguin}
shows the dependence of $\alpha$ on $H/\lambda$ for perfect metals
and a comparison to pairwise summation approximation, with the
dashed-line showing an empirical approximate formula
$\alpha_e=1/\cosh^{4/9}\left(\frac{12
\pi}{\sqrt{35}}\frac{H}{\lambda}\right)$ which could be useful for
practical purposes. Using this result we find
\begin{equation}
V_{\rm S}=\left(\frac{7 \pi^2 \sqrt{2}}{3072}\right)^{1/2}
\left(\frac{\hbar c}{\rho H^4}\right)^{1/2} \left(\frac{a_1
a_2}{H^2}\right)^{1/2}\left(\frac{H}{R}\right)^{3/4}
\alpha^{1/2},\label{VS-res-1}
\end{equation}
where $\rho$ is the mass density of the pinion. This result shows a
strong power law behavior at small values of $H$ followed by an
exponential decay at large $H$ whose length scale is set by
$\lambda$. Figure \ref{fig:derjaguin} shows the skipping velocity as
a function of the gap size, for different values of the radius and
wavelength. The typical values for $V_{\rm S}$, which correspond to
velocities that the system could transfer robustly, are remarkably
high: translated into angular velocity they are in the kHz region.
The strong dependence of the skipping velocity on the gap size can
be used to explore the parameter space and change the behavior of
the system. For example, one can reduce $V_{\rm S}$---move in the
vertical direction in the phase diagram of Fig.
\ref{fig:VRVP-combo}b---by increasing $H$, and thus switch the
system from the locked-in phase to a reverse gear, at constant rack
velocity. In other words, changing the separation provides a
continuous analogue of the clutch--gear system. The calculations
presented here have been based on the assumption of perfect metallic
boundaries, and one expects corrections for gap sizes smaller than
the plasma wavelength of the metals \cite{Reynaud}.

The value of the friction coefficient $\zeta$ is also instrumental
in determining the behavior of the system. While this quantity is
highly system-dependent in general, it is interesting to note that a
contribution to dissipation also comes from the interplay between
the electromagnetic fluctuations and the dielectric loss properties
of the two objects \cite{Pendry}.

In conclusion, we have proposed a design for a nano-scale rack and
pinion without contact by employing the quantum fluctuations of the
electromagnetic field. This design, and the corresponding variants
that could be readily conceived, might help towards making more
durable machine parts for small mechanical system.

\acknowledgements

This work was supported by EPSRC under Grant EP/E024076/1 (R.G.).

\end{document}